# A Simple and Effective Closed-Form GN Model Correction Formula Accounting for Signal Non-Gaussian Distribution

P. Poggiolini, G. Bosco, A. Carena, V. Curri, Y. Jiang, F. Forghieri

*Abstract*— **The GN model of non-linear fiber propagation has been shown to overestimate the variance of non-linearity due to the signal Gaussianity approximation, leading to maximum reach predictions for realistic optical systems which may be pessimistic by about 5% to 15%, depending on fiber type and system set-up. Analytical corrections have been proposed, which however substantially increase the model complexity. In this paper we provide a simple closed-form GN model correction formula, derived from the EGN model, which we show to be quite effective in correcting for the GN model tendency to overestimate non-linearity. The formula also permits to clearly identify the correction dependence on key system parameters, such as span length and loss.**

## I. INTRODUCTION

BUILDING on results from several similar prior modeling efforts [1]-[5] the GN model of non-linear propagation has been recently proposed as a practical tool for predicting the performance of uncompensated optical coherent transmission systems, in realistic scenarios [6]-[14]. A more extensive bibliography and a comprehensive model description are provided in [11], [14].

The GN model is characterized by remarkable simplicity, which was achieved thanks to several drastic approximations in its derivation [14]. Such approximations, however, inevitably cause errors in the estimation of non-linearity, or non-linear interference (NLI) noise.

The GN model accuracy has been the subject of recent investigations [14]-[19]. Interestingly, these studies have shown the incurred errors to be mostly related to one of the model approximations: the 'signal Gaussianity', which assumes that, over uncompensated links, the signal statistically behaves as Gaussian noise.

Specifically, [15] was the first paper to study in detail the inaccuracy incurred by the GN model when used to predict how NLI noise accumulates span-by-span along practical systems links. This simulative study showed that over the first few spans, where the signal is farther from Gaussian-distributed, the GN model strongly overestimates NLI noise power, up to several dB's. Such error then abates steadily along the link, but it is still significant at longer reaches, where a 1 to 2 dB NLI noise power overestimation can be seen, for typical systems.

Remarkable progress in the characterization of NLI accumulation was then made in [17] which succeeded in analytically removing the signal Gaussianity approximation for one of the main contributions to NLI (the cross-channel-modulation, or 'XPM'). The XPM analytical formulas have been used in [18] to generate various results which appear to be in general qualitative agreement with the simulative results of [15]. Such results also confirmed the dependence of non-linearity generation on the fourth moment of the signal constellation, that other groups found within different modeling approaches [20]-[21].

Even though the amount of NLI noise overestimation may be significant, the actual GN model error on the prediction of key system performance indicators, such as maximum reach or optimum launch power, is contained. Recent in-depth investigations [14], [15], [19] have shown that, when realistic system scenarios are considered, the GN model error on maximum system reach prediction (vs. simulations) is typically in the range 0.2-0.6 dB (5%-15%). One reason why these errors are relatively small is that the main system performance indicators have a low sensitivity to NLI power quantitative deviations: one dB error in NLI power estimation leads to only 1/3 dB error on either maximum reach or optimum launch power prediction [11], [14]. It should also be mentioned that, since the GN model errors are always biased towards overestimating NLI noise power in PM-QAM (polarization-multiplexed quadrature-amplitude-modulation) systems [15], [19], the GN model is always conservative, i.e., it never predicts a longer reach than simulations actually show.

The limited extent and conservative nature of these system performance prediction errors suggest that they could perhaps be dealt with through some heuristic correction. The 'incoherent GN model' [14] is an example of this. Its better accuracy is however due to two approximations canceling each other out by chance [14], [15]. A more rigorous solution, resting on sounder theoretical ground, is therefore desirable.

As mentioned, in [17] the authors analytically removed the Gaussianity assumption from the estimation of one of the NLI

Paper submitted on Feb. 26th 2014, revised July 22nd 2014. P. Poggiolini, G. Bosco, A. Carena, V. Curri, and Y. Jiang are with Dipartimento di Elettronica e Telecomunicazioni, Politecnico di Torino, Corso Duca degli Abruzzi 24, 10129, Torino, Italy, e-mail: pierluigi.poggiolini@polito.it; F. Forghieri is with CISCO Photonics, Via Santa Maria Molgora 48 C, 20871 Vimercate (MB), Italy, e-mail fforghie@cisco.com. This work was supported by CISCO Systems within a sponsored research agreement (SRA) contract.

noise components. We extended and generalized the procedure, to rigorously derive a complete 'enhanced' GN model (the 'EGN' model [19]) which addresses all NLI components with greatly improved accuracy vs. the GN model. However, although this approach is theoretically rigorous, the EGN model resulting complexity is substantially larger than that of the GN model, which can make its extensive practical use difficult [19].

In this paper we propose instead a very simple, closed-form correction to the GN model. It is fully justified on theoretical ground, since it is derived from the EGN model formulas. The GN model, together with the correction formula proposed here, actually provide a low-complexity approximation to the EGN model. Such approximation has limitations, which are fully discussed in the following, but already in its present form it effectively and rather accurately corrects for the GN-model bias towards NLI overestimation, without substantially increasing the GN model complexity. In addition, the availability of a closed-form GN model correction formula allows to clearly identify the correction dependence on key system parameters, such as span length and loss.

We carefully validate the GN model correction formula over a wide range of system scenarios. Besides serving this purpose, this validation campaign further confirms the excellent accuracy of the EGN model, which is compared with simulations and used as benchmark.

We also discuss the XPM approximation, as defined in [17], Eq. (25), which has been proposed as an overall NLI-estimation model (excluding single-channel non-linearity) incorporating the effects of signal non-Gaussianity. Our results indicate that the XPM approximation tends to underestimate NLI, in particular over low-dispersion fibers. We discuss the origin of such underestimation and the relationship of the XPM approximation with the closed-form GN model correction formula presented here.

The paper is organized as follows. In Sect. II we directly introduce the GN model correction formula and the resulting EGN model approximation. The details of its derivation are shown in Appendix A. In Sect. III we validate it by means of an extensive simulative NLI noise accumulation study. In Sect. IV we test it in the context of maximum system reach estimation. In Sect. V we point out the main parameter dependencies of the non-Gaussianity correction part of the approximate EGN model formula. In Sect. VI we discuss the paper main results. Conclusions follow.

## II. THE EGN MODEL APPROXIMATION

Throughout the paper we assume dual-polarization propagation, over realistic fibers with non-zero loss. The EGN model [1] approximation, whose derivation is reported in

---

[1] The EGN model is based on the Manakov equation, which accounts for the non-linear effect of one polarization on the other [22]. We use its simplified version consisting of the left-hand side of Eq. (12) in [22], which disregards polarization-mode dispersion (PMD). The linear effect of PMD is no longer a factor in modern coherent systems thanks to receiver digital signal processing (DSP). As for the non-linear impact of PMD, in [22] it was assessed to be very small or negligible in typical transmission links. Though PMD may have some

Appendix A, is shown in the following. Calling $G_{\text{NLI}}^{\text{EGN}}(f)$ the power spectral density (PSD) of NLI noise according to the EGN model [19], it is:

$$G_{\text{NLI}}^{\text{EGN}}(f) \approx G_{\text{NLI}}^{\text{GN}}(f) - G_{\text{corr}}$$
**Eq. 1**

where:

$$G_{\text{corr}} = \frac{80}{81} \Phi \frac{\gamma^2 \overline{L}_{\text{eff}}^2 P_{\text{ch}}^3 N_s}{R_s^2 \Delta f \pi \beta_2 \overline{L}_s} \text{HN}([N_{\text{ch}} - 1]/2)$$
**Eq. 2**

and $G_{\text{NLI}}^{\text{GN}}(f)$ is the NLI PSD according to the (coherent) GN model ([14], Eq. 2). The term $G_{\text{corr}}$ is a closed-form 'correction' which approximately corrects the GN model for the errors due to the signal Gaussianity assumption.

The meaning of the symbols is as follows:

- $f$ : optical frequency (THz), with $f = 0$ conventionally being the center frequency of the center channel
- $\alpha$ : optical field fiber loss (1/km), such that the optical *field* attenuates as $e^{-\alpha z}$; note that the optical *power* attenuates as $e^{-2\alpha z}$
- $\beta_2$ : dispersion coefficient (ps²/km)
- $\gamma$ : fiber non-linearity coefficient, 1/(W km)
- $\overline{L}_s$ : average span length (km)
- $\overline{L}_{\text{eff}}$ : average span effective length (km), with span effective length defined as $L_{\text{eff}} = (1 - e^{-2\alpha L_s})/2\alpha$
- $N_s$ : total number of spans in the link
- $N_{\text{ch}}$ : total number of channels in the system
- $P_{\text{ch}}$ : the launch power per channel (W)
- $\Delta f$ : channel spacing (THz)
- $R_s$ : symbol rate (TBaud)

The specified units ensure consistency if used to express the parameters in Eq. 2. In addition, $\text{HN}(N)$ is the harmonic number series, defined as:

$$\text{HN}(N) = \sum_{n=1}^{N}(1/n)$$
**Eq. 3**

Finally, $\Phi$ is a constant that depends on the modulation format (see [19] and App. A of this paper). Its values are: 1, 17/25 and 13/21 for PM-QPSK, PM-16QAM and PM-64QAM, respectively.

---

impact on NLI, we consider neglecting it a reasonable approximation, for the purpose of achieving manageable analytical modeling.

Eq. 2 assumes that all channels are identical and equally spaced. This assumption can be removed but this topic will not be dealt with in this paper. It also assumes that channels are single-carrier type (neither OFDM nor massively multi-subcarrier).

Eq. 2 assumes that the same type of fiber is used in all spans. Spans can be of different length, though: Eq. 2 uses the average span length $\bar{L}_s$ and the average span effective length $\bar{L}_{eff}$. Accuracy is quite good for links having all individual span lengths within $\bar{L}_s \pm 15\%$. Caution should be used for larger deviations.

Eq. 2 also assumes lumped amplification, exactly compensating for the loss of the preceding span. Regarding the use of Eq. 1 with Raman-amplified systems, see discussion in Sect. VI.

Eq. 2 has the following further limitations.

- $G_{corr}$ approximately corrects the cross-channel interference contributions to NLI. It does not correct the single-channel interference (SCI, see App. A and [11], [19]). Therefore, the overall Eq. 1 is increasingly more accurate as the number of channels is increased, whereas for a single-channel system it coincides with the standard GN model. A fully analytical correction for SCI is available as part of the EGN model [19], but currently not in simple closed-form.

- $G_{corr}$ is asymptotic in the number of spans. As a result, its accuracy improves as the number of spans grows. The speed of the asymptotic convergence depends on the number of channels and on fiber dispersion (see Sect. III).

- $G_{corr}$ is derived assuming ideally rectangular channel spectra. If spectra have a significantly different shape (such as sinc-shaped), some accuracy may be lost.

- $G_{corr}$ is calculated at $f = 0$ and then it is assumed to be frequency-flat.

We point out that a less approximate expression for $G_{corr}$ than Eq. 2 is provided in App. A, as Eq. 16 and Eq. 18 combined. Its convergence vs. $N_s$ is faster and more accurate. It is however not closed-form as it requires a final one-dimensional numerical integration. In this paper we concentrate on the simpler Eq. 2.

## III. VALIDATION OF $G_{corr}$

As pointed out, $G_{corr}$ does not correct the single-channel interference (SCI) contribution to non-linearity. Therefore, we focus its specific validation effort on the other two NLI components, XCI and MCI (cross- and multi-channel interference [11]), which we call together XMCI, for brevity. In other words, XMCI is the total NLI, except for SCI which is removed in the following.

As a consequence, a straightforward choice for the quantity to focus on for model validation could be:

$$P_{XMCI} = \int_{-R_s/2}^{R_s/2} G_{XMCI}(f)\, df$$

**Eq. 4**

It represents the total XMCI noise spectrally located within the center WDM channel. However, $P_{XMCI}$ depends on the signal launch power. Specifically, it is proportional to $P_{ch}^3$. If $P_{XMCI}$ is normalized with respect to $P_{ch}^3$, then the resulting quantity does not change vs. the launch power and becomes a power-independent characterization of the XMCI behavior of the link. We therefore decided to concentrate on the normalized quantity $\eta_{XMCI}$, defined simply as:

$$\eta_{XMCI} = \frac{P_{XMCI}}{P_{ch}^3}.$$

**Eq. 5**

We estimated $\eta_{XMCI}$ in three ways:

1. through accurate computer simulations;
2. calculating $G_{XMCI}(f)$ in Eq. 4 using the EGN model formulas for the XMCI power spectral density provided in [19];
3. approximating $G_{XMCI}(f)$ in Eq. 4 using the EGN model approximation Eq. 1 with the single-channel non-linearity (SCI) contribution removed from the GN model term:

$$G_{XMCI}(f) \approx G_{NLI}^{GN}(f) - G_{corr} - G_{SCI}^{GN}(f)$$

**Eq. 6**

For comparison, we also considered XPM as defined in [17], Eq. (25), which was proposed as a possible alternative model for overall NLI estimation (excluding single-channel non-linearity), which incorporates the effects of signal non-Gaussianity too.

Regarding the computer simulations, the same simulation software, simulation techniques and general system set-up described in [14], Sect. V, were used. The main details are reported in the following.

The fiber simulation algorithm is based on the standard split-step integration technique. The simulated systems symbol rate was $R_s = 32$ GBaud, with raised-cosine signal PSD and roll-off 0.05. The channel spacing was 33.6 GHz. The launch-power was -3 dBm per channel. Note that the quantity $\eta_{XMCI}$ is defined so as to be launch-power independent but nonetheless we redid some of the simulations at both -6 and 0 dBm to check whether any changes could be seen. We found no significant difference.

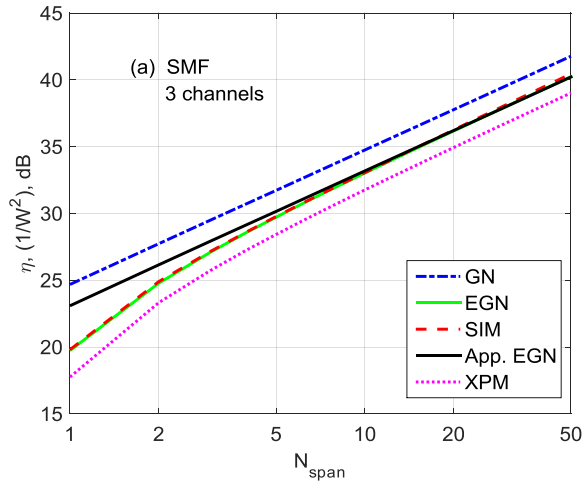

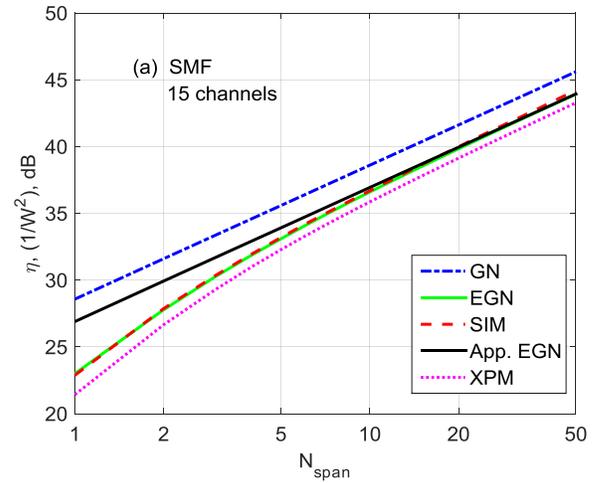

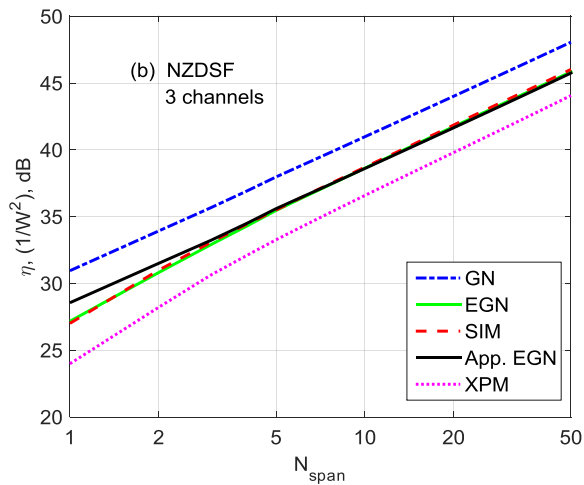

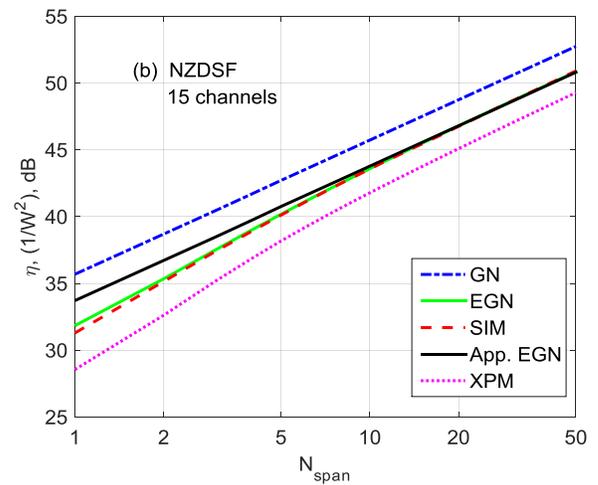

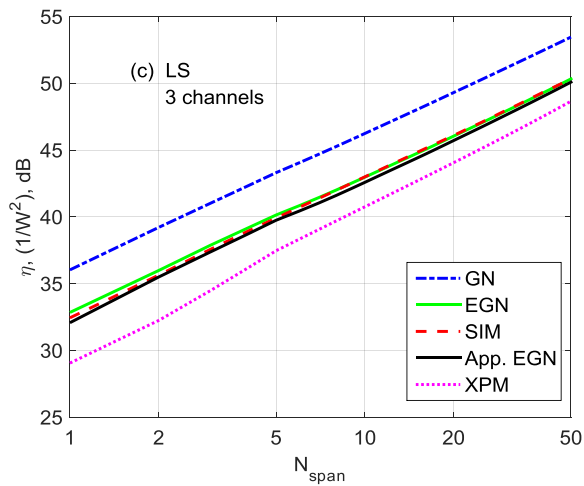

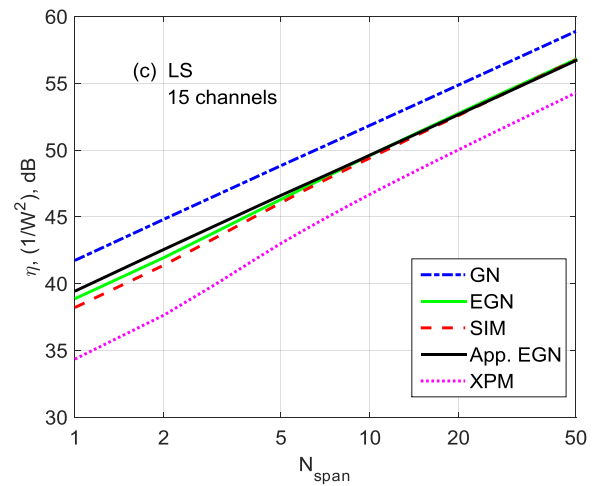

Fig. 1: Plot of the normalized combined cross- and multi-channel non-linearity noise power coefficient $\eta_{XMCI}$ affecting the center channel (in dB referred to $1 \cdot W^{-2}$), vs. number of spans in the link. Single-channel effects (SCI) are completely removed from all curves. System data: 3 PM-QPSK channels, 32 GBaud, roll-off 0.05, span length 100 km, channel spacing 33.6 GHz. The 'app. EGN' curve is generated using Eq. 6.

Fig. 2: Plot of the normalized combined cross- and multi-channel non-linearity noise power coefficient $\eta_{XMCI}$ affecting the center channel (in dB referred to $1 \cdot W^{-2}$), vs. number of spans in the link. Single-channel effects (SCI) are completely removed from all curves. System data: 15 PM-QPSK channels, 32 GBaud, roll-off 0.05, span length 100 km, channel spacing 33.6 GHz. The 'app. EGN' curve is generated using Eq. 6.

The tested fibers were: standard single-mode (SMF) with $D = 16.7$ ps/(nm·km), $\gamma = 1.3$ (W·km)$^{-1}$; non-zero dispersion-shifted fiber (NZDSF, similar to OFS's TrueWave RS), with $D = 3.8$ ps/(nm·km), $\gamma = 1.5$ (W·km)$^{-1}$; negative non-zero dispersion-shifted fiber (which we call "LS" because it is similar to Corning's LS fiber) with $D = -1.8$ ps/(nm·km), $\gamma = 2.2$ (W·km)$^{-1}$. The span length was $L_s = 100$ km and loss was $\alpha_{dB} = 0.22$ dB/km for all fibers.

To remove SCI, we ran a single-channel simulation and recorded the optical signal at the receiver (Rx). This signal was then subtracted from that of the WDM simulations. The total variance of the residual signal was measured and used to calculate the simulative $\eta_{XMCI}$ estimate.

The Rx compensated statically for polarization rotation and applied an ideal matched filter. No dynamic equalizer was used, to avoid any possible effect of the equalizer adaptivity on XMCI estimation. The simulation was completely noiseless: neither ASE noise, nor any other types of noise, such as Rx electrical noise, were present.

A first set of results is plotted in Fig. 1-Fig. 2. The quantity $\eta$, whose units are $1/W^2$, is reported [2] in dB. We chose PM-QPSK as modulation format because the strength of the non Gaussianity correction is maximum, since its coefficient $\Phi$ in Eq. 2 is the largest among QAM formats. We show 3-channel systems in Fig. 1 and 15-channel systems in Fig. 2. The reason for choosing these channel numbers is that it was the largest channel number *range* that we could cover through simulations. We also have intermediate sets run at 5 and 9 channels, not shown here both for brevity and because their results are qualitatively very similar to those reported here.

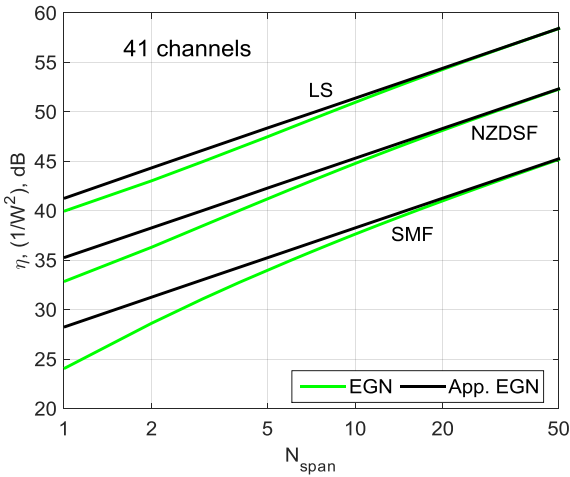

Fig. 3: Plot of the normalized combined cross- and multi-channel non-linearity noise power coefficient $\eta_{XMCI}$ affecting the center channel (in dB referred to 1·W$^{-2}$), vs. number of spans in the link. Single-channel effects (SCI) are completely removed from all curves. System data: 41 PM-QPSK channels, 32 GBaud, roll-off 0.05, span length 100 km, channel spacing 33.6 GHz. The 'app. EGN' curve is generated using Eq. 6.

[2] All the plots in Figs. 1-6 display the quantity $\eta$, whose units are W$^{-2}$, in dB referred to unity in the specified units, that is, vs. 1·W$^{-2}$.

A common feature of all these plots is that the EGN model shows very good accuracy in estimating XMCI, throughout all system configurations, confirming the findings in [19] and confirming itself as a reliable reference benchmark.

The GN model always overestimates XMCI, along the lines of what was found in [15]-[19]. The extent of the overestimation depends on fiber dispersion and behaves in a peculiar way. The higher the dispersion, the greater the error for low span count, but the lower for high span count. In fact, at 15 channels, SMF is the fiber for which the GN model shows both the highest 1-span error (5.7 dB) and the lowest 50-span error (1.4 dB).

The XPM approximation of [17], Eq. (25), underestimates non-linearity in the examples shown in Fig. 1-Fig. 2. Interestingly, we found that this error does not derive from the non-Gaussianity correction term present in the XPM approximation formula (called $\chi_2$ in [17]), which is quantitatively close to $G_{corr}$. Rather, it is caused by the GN model-like contribution (called $\chi_1$ in [17]) which is underestimated, especially over NZDSF and LS fibers. This in turn derives from the assumption made in [17] that XPM is the dominant component to NLI, so that the other components are discarded. At least in these examples, the discarded contributions (part of XCI and all of MCI) are relevant and cannot be neglected[3].

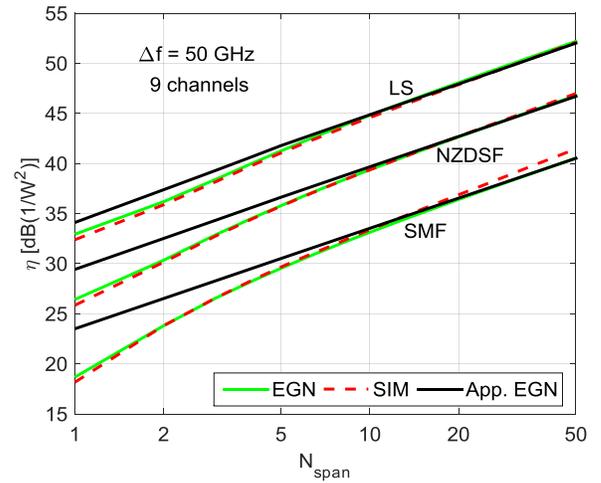

Fig. 4: Plot of the normalized combined cross- and multi-channel non-linearity noise power coefficient $\eta_{XMCI}$ affecting the center channel (in dB referred to 1·W$^{-2}$), vs. number of spans in the link. Single-channel effects (SCI) are completely removed from all curves. System data: 9 PM-QPSK channels, 32 GBaud, roll-off 0.05, span length 100 km, channel spacing 50 GHz. The 'app. EGN' curve is generated using Eq. 6.

The approximate EGN model Eq. 6, relying on the simple correction Eq. 2, is quite effective with all fibers, showing good convergence towards the exact EGN model curve and vs. simulations, as the number of spans grows. As a result of its asymptotic behavior, Eq. 2 only partially recovers the GN model at low span count. On the other hand, at span counts that are typically of interest for maximum reach predictions, its accuracy is good. The error vs. either simulations or the EGN model curve is less than 0.4 dB in the whole range 10-50 spans, for any number of channels among 3,5,9,15 (5 and 9 not shown), for all three analyzed fibers. It stays below 0.7 dB even down to only 5 spans, across all cases.

[3] More in depth, XPM as defined in Eq. (25) of [17] is equivalent to evaluating the EGN model limited to only the 'X1' integration regions shown in reference [19], Fig. 7. All other integration regions are neglected by the XPM approximation. It turns out that while the non-Gaussianity correction outside of the X1 regions can often be neglected, this is not the case for the GN model contribution. The latter must typically be computed over some of the other regions too (see [19], Sects. 4 and 5, for more details).

### A.  Higher channel count

We wanted to check whether a similarly reliable behavior was maintained at a substantially higher channel count. We looked at 41 channels where, however, we could not run benchmark simulations because of the excessive required computation time. The check is therefore made towards the EGN model curve alone. Fig. 3 shows the $\eta_{\text{XMCI}}$ results for the three reference fibers. The very good asymptotic convergence of Eq. 2 towards the EGN model is confirmed even at this substantially higher channel count.

### B.  Larger channel spacing

To see whether Eq. 2 held up at larger channel spacing, we ran checks at 50 GHz spacing, with 9 channels. Fig. 4 shows that Eq. 2 is asymptotically accurate at this spacing as well, on all three reference fibers.

### C.  Shorter span lengths

We also ran a set of checks at a substantially shorter span length $L_s$ =60 km. Fig. 5 shows a quite good overall convergence of the approximate EGN estimate, even at relatively low number of spans.

### D.  PM-16QAM transmission

With PM-16QAM, the GN model correction becomes weaker, as the coefficient $\Phi$ in Eq. 2 shrinks from 1 (for PM-QPSK) to 17/25. Nonetheless, as shown both in [19] and in the next section of this paper, PM-16QAM maximum reach prediction is improved in a non-negligible way by correcting the GN model. Therefore, it is desirable that the approximate correction formula Eq. 2 performs well for this format, too. The detailed $\eta_{\text{XMCI}}$ vs. number of spans result obtained for 9-channel systems, with spacing 33.6 GHz, is shown in Fig. 6. Simulations and the EGN model prediction are in excellent agreement. The accuracy of the asymptotic formula is very good, too.

## IV.  SYSTEM PERFORMANCE PREDICTION

The main declared goal of many of the recent modeling efforts has been that of providing a practical tool for realistic system performance prediction. In this section we present a comparison of the accuracy of the GN model and of the approximate EGN model of Eq. 1 in predicting maximum system reach in some typical scenarios.

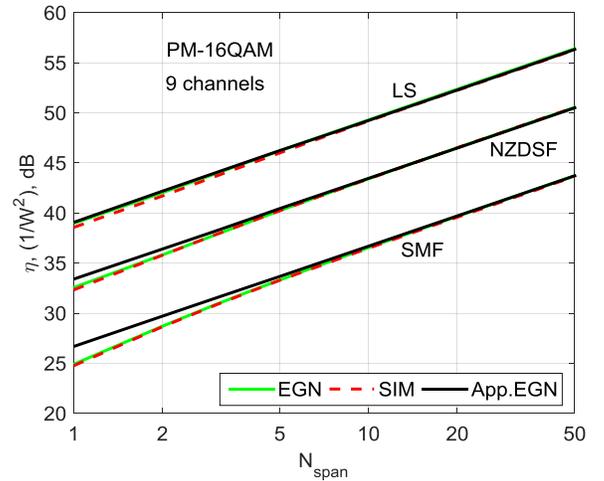

Fig. 6: Plot of the normalized combined cross- and multi-channel non-linearity noise power coefficient $\eta_{\text{XMCI}}$ affecting the center channel (in dB referred to $1 \cdot \text{W}^{-2}$), vs. number of spans in the link. Single-channel effects (SCI) are completely removed from all curves. System data: 9 PM-16QAM channels, 32 GBaud, roll-off 0.05, span length 100 km, channel spacing 33.6 GHz. The 'app. EGN' curve is generated using Eq. 6.

Note that the EGN model accuracy in predicting system maximum reach was tested in [19], Sect. 6, and found to be excellent, at least in the tested cases, which are the same as those addressed there. Specifically, they are 15-channel PM-QPSK and PM-16QAM systems, running at 32 GBaud. We considered the following channel spacings: 33.6, 35, 40, 45 and 50 GHz. The spectrum was root-raised-cosine with roll-off 0.05. The target BERs were $1.7 \cdot 10^{-3}$ and $2 \cdot 10^{-3}$ respectively, found by assuming a $1 \cdot 10^{-2}$ FEC threshold, decreased by 2 dB of realistic OSNR system margin. EDFA amplification was assumed, with 5 dB noise figure. Note that, differently from

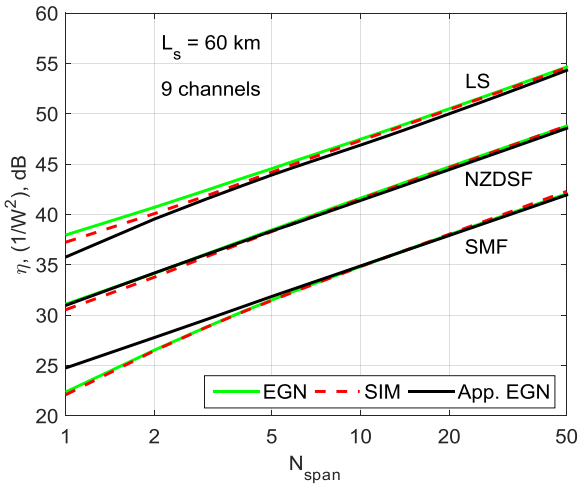

Fig. 5: Plot of the normalized combined cross- and multi-channel non-linearity noise power coefficient $\eta_{\text{XMCI}}$ affecting the center channel (in dB referred to $1 \cdot \text{W}^{-2}$), vs. number of spans in the link. Single-channel effects (SCI) are completely removed from all curves. System data: 9 PM-QPSK channels, 32 GBaud, roll-off 0.05, span length 60 km, channel spacing 33.6 GHz. The 'app. EGN' curve is generated using Eq. 6.

Fig. 1, single-channel non-linear effects were *not* removed from the simulations. The considered fibers were: SMF, NZDSF and LS, with the same parameters as before, except for SMF whose loss was set to $\alpha_{dB} = 0.2$ dB/km. In addition, we considered pure-silica-core fiber (PSCF) with the following parameters: $D =$20.1 ps/(nm·km), $\gamma =$0.8 (W·km)$^{-1}$, $\alpha_{dB} =$0.17 dB/km.

We point out that we did not assume that the spectrum of NLI was flat, i.e., we did not use the so-called 'white-noise approximation'. We did take into account its actual shape when estimating the system maximum reach, either based on the GN model alone or based on Eq. 1. Note though that, as pointed out in Sect. II, the approximate correction Eq. 2 is assumed frequency-independent. We also point out that the simulative results of this section are found by adding all ASE noise at the end of the link, rather than in-line. The reason for this is that here we want to validate an approximate model that neglects the interaction of in-line ASE noise with non-linearity. Not plotted (for the sake of clarity), the simulative data points with in-line ASE noise are on average about 0.15 dB lower (on $N_{span}$) for PM-QPSK. The effect on PM-16QAM is instead negligible, because PM-16QAM requires a much higher OSNR at the receiver and hence much less ASE noise propagates along the link than for PM-QPSK.

Fig. 7 shows a plot of maximum system reach vs. channel spacing. The GN model underestimates the maximum reach by 0.3–0.6 dB over PSCF, SMF and NZSDF, and up to 0.8 dB over the ultra-low dispersion LS, in agreement with [14], [15] and [19]. These results are also in line with the general picture that emerges from Fig. 1-Fig. 2 and Fig. 4-Fig. 6, when taking into account that an error of $x$ dB in the estimation of NLI power leads to an error of about $x/3$ dB in maximum reach estimation [14].

With all fibers, the approximate EGN model Eq. 1 is quite effective and for low frequency spacing (33.6 and 35 GHz) the predictions based on it come within a quite small error range [−0.2, 0] dB across all scenarios. The error range widens slightly to [−0.4, −0.1] dB for the larger frequency spacings. Since Eq. 2 does not appear to lose accuracy at 50 GHz (see Fig. 4) we do not think that the somewhat larger error can be ascribed to it. Rather, it could be ascribed to the fact that Eq. 1 neglects the non-Gaussianity correction for single-channel non-linearity (SCI). This means SCI is overestimated, leading to a pessimistic maximum reach prediction. The impact of such error is greater at larger channel spacings because single-channel effects have a greater relative impact at larger spacings than for quasi-Nyquist spacing.

On the other hand, if the number of channel $N_{ch}$ increases, the maximum reach error decreases, because SCI is a fixed quantity whereas XMCI grows vs. $N_{ch}$. For example, at 50 GHz spacing, we found that the increase in XMCI when going from $N_{ch}$ =15 to $N_{ch}$ = 41 is 1.3 dB, for PM-QPSK over SMF at 30 spans. It is similar, (1.4 dB) over NZDSF at 15 spans.

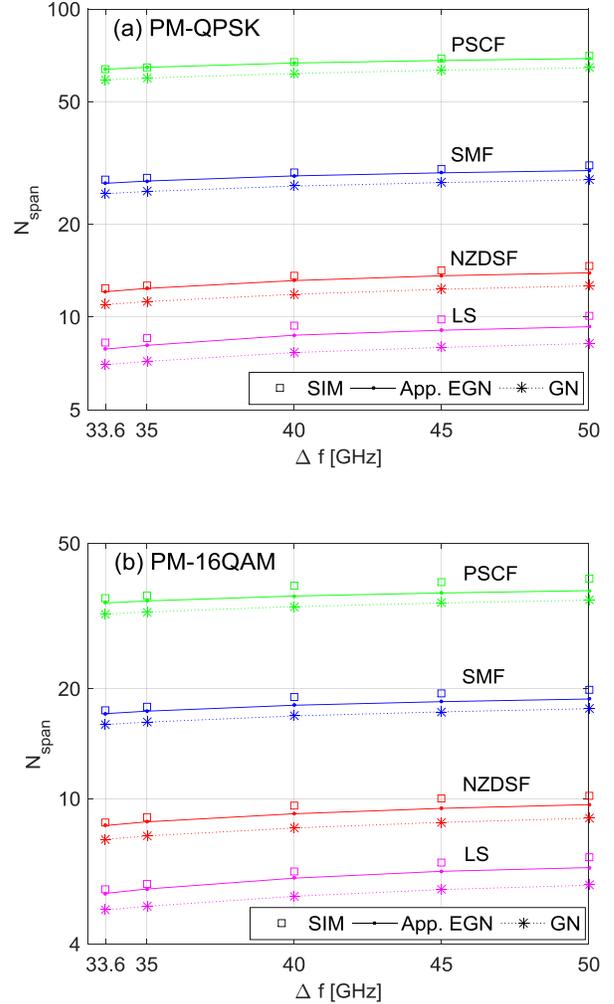

Fig. 7: Plot of maximum system reach for 15-channel PM-QPSK and PM-16QAM systems at 32 GBaud, roll-off 0.05, vs. channel spacing $\Delta f$, over four different fiber types: PSCF, SMF, NZDSF and LS. The span length is 120 km for PM-QPSK and 85 km for PM-16QAM. The 'app. EGN' curve is generated using Eq. 1.

## V. PARAMETER DEPENDENCIES OF THE APPROXIMATE EGN MODEL

Eq. 1 can be made fully closed-form by substituting $G_{NLI}^{GN}(f)$ with one of the GN model approximations described for instance in [11]. We discuss here a specific example, that of ideal Nyquist WDM transmission with all-identical spans ([11], Eq. 15), for the sole purpose of pointing out certain parameter dependencies of the resulting formula. NLI is evaluated at the center of the center channel ($f = 0$).

Due to the approximations used in [11] to derive Eq. 15 there, to combine such formula with Eq. 2 meaningfully we have to assume that for all the spans in the link the following approximation is accurate enough:

$$\left[1 - \exp(-2\alpha L_s)\exp(j\varphi)\right] \approx 1$$

where $L_s$ is the span length of any single span and $\varphi$ has a complex expression (see App. A, Eq. 12) and in general can vary over $[0, 2\pi]$. Therefore, the remarks made in the following are valid only if the loss of all of the spans in the link is greater than approximately 10 dB. If so, we can then write:

$$G_{NLI}^{EGN}(0) \approx G_{NLI}^{GN}(0) - G_{corr} \approx \frac{4}{27} \frac{\gamma^2 P_{ch}^3 N_s}{R_s^3 \pi \beta_2 \alpha} \cdot$$
$$\left[ N_s^\varepsilon \, \mathrm{asinh}\left(\tfrac{1}{4} \beta_2 \alpha^{-1} \pi^2 N_{ch}^2 R_s^2\right) - \frac{10}{3} \Phi \frac{1}{2\alpha L_s} \mathrm{HN}\left([N_{ch}-1]/2\right) \right]$$

**Eq. 7**

where 'asinh' is the hyperbolic arcsine. The symbol $\varepsilon$ is the NLI noise coherent accumulation exponent, with typically $\varepsilon \ll 1$ [11]. The first term in square brackets derives from $G_{NLI}^{GN}(f)$ whereas the second term stems from the non-Gaussianity correction $G_{corr}$ of Eq. 2. The formula shows that these two terms have important common dependencies, which appear as common factors outside the square brackets, such as $\gamma^2$, $P_{ch}^3$ and $1/\beta_2$. Note that the presence of $\beta_2$ in the asinh function has little effect because asinh is a log-like slowly increasing function.

From Eq. 7 one can directly derive the relative strength of the non-Gaussianity correction $G_{corr}$ vs. the GN model contribution $G_{NLI}^{GN}(0)$, which can be written as:

$$G_{corr} \Big/ G_{NLI}^{GN}(0) \approx \frac{\dfrac{10}{3} \Phi \dfrac{1}{2\alpha L_s} \mathrm{HN}\left([N_{ch}-1]/2\right)}{N_s^\varepsilon \, \mathrm{asinh}\left(\tfrac{1}{4} \beta_2 \alpha^{-1} \pi^2 N_{ch}^2 R_s^2\right)}$$

**Eq. 8**

One interesting aspect is that this ratio is inversely proportional to the span length $L_s$. It is also inversely proportional to the span loss coefficient $\alpha$, though approximately, because $\alpha$ is also present in the argument of the asinh function. However, the log-like nature of asinh dampens its variations so that the $1/\alpha$ factor at the numerator of Eq. 8 sets the prevailing trend for typical values of the other parameters.

Neglecting the asinh variation, then it appears that Eq. 8 is inversely proportional to the overall span loss, expressed as $1/(2\alpha L_s)$. In other words, the non-Gaussianity correction has more impact over low span-loss systems. Conversely, it tends to vanish for high-loss spans. This is in agreement with what simulatively or numerically predicted in [17]-[19], but here this dependence stands out analytically. Once again, though, note that the above formula is accurate only as long as span loss is greater than about 10 dB, i.e., for $1/(2\alpha L_s) \leq 0.43$.

## VI. DISCUSSION

As mentioned in the previous section, and also as found elsewhere [14], [15], [19], the inaccuracy incurred by the coherent GN model ([14], Eq. 2) in system maximum reach assessment, vs. simulations results, is typically about 0.3-0.6 dB, reaching 1 dB only on ultra-low dispersion fibers and/or when using very short spans. In addition, such error is always conservative, i.e., it is biased vs. predicting a shorter reach (see Fig. 7). Depending on the specific use, this performance may or may not be adequate.

If high-accuracy span-by-span NLI estimation is needed, the EGN model [19], developed by generalizing the approach of [17], proves very effective, as clearly confirmed in Fig. 1-Fig. 2 and Fig. 4-Fig. 6. The EGN model is however rather complex and computationally heavy and it is difficult to consider it a realistic alternative for agile system studies.

Given its closed-form and great simplicity, Eq. 2 therefore represents a potentially quite helpful tool for achieving a system maximum reach prediction accuracy close to that of the EGN model, with essentially the same complexity of the GN model for which efficient computation techniques have been proposed (see for example [11], [23]-[25]). Its asymptotic convergence appears to be very robust across many system variants and configurations.

Raman amplification is currently drawing substantial interest, especially in the context of terrestrial systems with long span lengths. Although derived for lumped amplification, Eq. 1 can be used to address these systems too, as long as non-linearity generation is scarcely affected by Raman. This is the case for backward-pumped Raman-amplified long spans, where span loss is on the order of 20 dB or more, and the on-off Raman gain is substantially lower than the total span loss, by at least 6 dB as an indicative figure. If so, the signal power towards the end of the span stays well below the level at the beginning of the span and its contribution to the total span non-linearity is small or negligible. The effect of Raman can then be completely ignored in Eq. 1. From a system point of view, Raman would only contribute to lowering the span-equivalent noise figure.

## VII. CONCLUSION

In conclusion, we have presented a compact, closed-form simple correction to the GN model, based on an approximation of the very accurate but complex EGN model [19].

The formula improves the GN model accuracy by suppressing most of its tendency to overestimate non-linearity. We have provided quite extensive validation. Albeit approximate, the formula is firmly based on theory and it proves quite effective.

Among its limitations, which could be addressed in the future to improve it, is the neglect of correcting single-channel non-linearity overestimation due to signal non-Gaussianity. This limitation has however relatively little impact in the context of WDM systems with a significant number of channels.

In summary, already in this form, the GN model correction formula provides a very effective tool that substantially improves the overall accuracy of the GN model in predicting realistic WDM system performance without significantly increasing its computational complexity.

## APPENDIX A: DERIVATION OF EQ. 2

In [19] we proposed the EGN model, which consists of a complete set of analytical formulas for all types of NLI (SCI, XCI and MCI). We derived them by generalizing the approach proposed in [17] to remove the signal Gaussianity assumption from the GN model calculations.

The EGN model can be compactly written similarly to Eq. 1:

$$G_{\text{NLI}}^{\text{EGN}}(f) = G_{\text{NLI}}^{\text{GN}}(f) - G_{\text{corr}}^{\text{ex}}(f)$$

**Eq. 9**

where $G_{\text{corr}}^{\text{ex}}(f)$ is a correction term to the GN model estimate of the PSD of NLI, $G_{\text{NLI}}^{\text{GN}}(f)$. The superscript 'ex' stands for 'exact' and is meant to distinguish it from the approximate correction $G_{\text{corr}}$ shown as Eq. 2. In the following, we show how to derive Eq. 2 from (formally) Eq. 9.

First of all, we impose that the term $G_{\text{NLI}}^{\text{GN}}(f)$ contains all of NLI (SCI, XCI and MCI). We stress the fact that neglecting parts of either XCI or MCI in the GN model term $G_{\text{NLI}}^{\text{GN}}(f)$ may lead to quite substantial error, as it is the case for instance for the XPM approximation [17] discussed in Sect. III (see Fig. 1). Closed-form approximations or ways to efficiently compute $G_{\text{NLI}}^{\text{GN}}(f)$ can be found in [11], [23]-[25] and will not be dealt with here.

The term $G_{\text{corr}}^{\text{ex}}(f)$ is more complex than $G_{\text{NLI}}^{\text{GN}}(f)$. To reduce it to the simple closed-form $G_{\text{corr}}$, several assumptions and approximations are necessary. First, we decided to neglect SCI in $G_{\text{corr}}$, because the exact SCI formulas appeared hard to reduce to closed-form. Hence SCI is going to be overestimated in Eq. 1, but in dense WDM systems, operating at high channel count, the majority of NLI comes from cross-channel effects and the error on SCI tends to become unimportant, as it is also shown by the system maximum reach results of Fig. 7.

Then, we studied the many XCI and MCI correction terms appearing in $G_{\text{corr}}^{\text{ex}}(f)$ and found that the dominant ones are just those whose integration domains straddle the axes of the $[f_1, f_2]$ plane, that is the domains of type 'X1' that appear in [19], Fig. 7. In essence, while in $G_{\text{NLI}}^{\text{GN}}(f)$ both XCI and MCI must be included, MCI needs not be corrected, as well as some parts of XCI, because their correction is either zero or is small. This circumstance has been extensively double-checked and confirmed by the many comparisons with the exact EGN model results obtained in a variety of system configurations, shown in Figs. 1 - 6 in this paper. Dropping all these lesser correction

terms, the following approximation to $G_{\text{corr}}^{\text{ex}}(f)$ is found:

$$G_{\text{corr}}^{\text{ex}}(f) \approx \sum_{n_{\text{ch}} \in \mathcal{N}_{\text{ch}}} \Phi \frac{80}{81} R_s^2 \gamma^2 P_{\text{ch}}^3 \int_{-\infty}^{\infty} df_1 \int_{-\infty}^{\infty} df_2 \int_{-\infty}^{\infty} df_3$$

$$\left| s_{\text{CUT}}(f_1) \right|^2 s_{\text{INT}_{n_{\text{ch}}}}(f_2) s_{\text{INT}_{n_{\text{ch}}}}^*(f_3) s_{\text{INT}_{n_{\text{ch}}}}^*(f_1 + f_2 - f)$$

$$s_{\text{INT}_{n_{\text{ch}}}}(f_1 + f_3 - f) \mu(f_1, f_2, f) \mu^*(f_1, f_3, f)$$

**Eq. 10**

The various quantities appearing in Eq. 10 are as follows. First,

$$\Phi = 2 - \mathbf{E}\left\{ |a_x|^4 + |a_y|^4 \right\} \Big/ \left[ \mathbf{E}\left\{ |a_x|^2 + |a_y|^2 \right\} \right]^2$$

**Eq. 11**

where $a_x$, $a_y$ are the random variables which represent the transmitted symbols over the two polarizations $\hat{x}$ and $\hat{y}$, and $\mathbf{E}\{\cdot\}$ is the statistical expectation operator. Then, the quantity $\mu$ was defined in [19] as the 'link function' and its expression, under the system assumptions listed in Sect. II, is:

$$\mu(f_1, f_2, f) = \sum_{n_s=1}^{N_s} \frac{1 - e^{-2\alpha L_s^{n_s}} e^{j q (f_1 - f)(f_2 - f) L_s^{n_s}}}{2\alpha - jq(f_1 - f)(f_2 - f)} e^{j(f_1 - f)(f_2 - f)qL_{\text{acc}}^{(n_s - 1)}}$$

**Eq. 12**

where $L_s^{n_s}$ is the length of the $n_s$-th span, $L_{\text{acc}}^{n_s} = \sum_{k=1}^{n_s} L_s^k$ is the accumulated length of the first $n_s$ spans, with $L_{\text{acc}}^0 = 0$, and $q = 4\pi^2 \beta_2$. The set $\mathcal{N}_{\text{ch}}$ contains all the indices $n_{\text{ch}}$ labeling the interfering channels (INT) present in the WDM system. From here on we assume the channel under test (CUT) to have index $n_{\text{ch}} = 0$ and the INT channels to have indices:

$$\mathcal{N}_{\text{ch}} = -(N_{\text{ch}} - 1)/2, \dots -1, 1, \dots, (N_{\text{ch}} - 1)/2$$

**Eq. 13**

where $N_{\text{ch}}$ is the total number of WDM channels (assumed odd).

The functions $s_{\text{CUT}}(f)$, $s_{\text{INT}_{n_{\text{ch}}}}(f)$ are the Fourier transforms of the pulses used by the channel under test (CUT) and by the $n_{\text{ch}}$-th interfering (INT) channel. The CUT is centered at $f = 0$ while the $n_{\text{ch}}$-th INT channel is located at $f = n_{\text{ch}} \cdot \Delta f$.

It is interesting to remark that Eq. 10 is similar to $\chi_2$ in Eq. (25) of [17]. However the GN model-like $\chi_1$ part in the same Eq. (25) of [17] is quite different than $G_{\text{NLI}}^{\text{GN}}(f)$ in Eq. 9) because of the choice in [17] to drop MCI and parts of XCI from the GN model contribution.

As a simplifying assumption, we assume all pulses to have rectangular Fourier transforms with bandwidth $R_s$. We set their flat-top value equal to $1/R_s$. Note that if so, then the channel power is given by:

$$P_{ch} = \mathbf{E}\left\{|a_x|^2 + |a_y|^2\right\}$$

**Eq. 14**

As another necessary approximation to achieve a simple closed-form result, we assume that $G_{corr}^{ex}(f)$ is approximately 'flat', i.e., frequency-independent, over the CUT bandwidth. Therefore we focus on calculating it at the center of the CUT, i.e., we focus on $G_{corr}^{ex}(0)$. As a result, we get:

$$G_{corr}^{ex}(0) \approx \sum_{n_{ch} \in \mathcal{N}_{ch}} \Phi \frac{80}{81} R_s^{-4} \gamma^2 P_{ch}^3$$
$$\int_{-R_s/2}^{+R_s/2} df_1 \int_{n_{ch}\Delta f - R_s/2}^{n_{ch}\Delta f + R_s/2} df_2 \int_{n_{ch}\Delta f - R_s/2}^{n_{ch}\Delta f + R_s/2} df_3 \; \mu(f_1, f_2, 0)\, \mu^*(f_1, f_3, 0)$$

**Eq. 15**

where we have also applied a further slight approximation in the domain of integration, consisting in replacing the lozenge-shaped X1 domains of [19], Fig. 7, with the square domains tightly inscribing them. This permits to formally remove the rectangular pulse spectra $s_{CUT}(f)$, $s_{INT_{n_{ch}}}(f)$ from the integrand, allowing to obtain Eq. 15 from Eq. 10. Inspection of Eq. 15 reveals that it can be exactly re-written as:

$$G_{corr}^{ex}(0) \approx \sum_{n_{ch} \in \mathcal{N}_{ch}} \Phi \frac{80}{81} R_s^{-4} \gamma^2 P_{ch}^3 \int_{-R_s/2}^{+R_s/2} |\zeta_{n_{ch}}(f_1)|^2 \, df_1$$

**Eq. 16**

where:

$$\zeta_{n_{ch}}(f_1) = \int_{n_{ch}\Delta f - R_s/2}^{n_{ch}\Delta f + R_s/2} \mu(f_1, f_2, 0) \, df_2$$

**Eq. 17**

We therefore concentrate on evaluating $\zeta_{n_{ch}}(f_1)$. Remarkably, Eq. 17) can be integrated analytically, albeit in terms of special functions.

$$\zeta_{n_{ch}}(f_1) = \frac{j}{q\, f_1} \Big\{ \ln\left(2\alpha - jq f_1\left[n_{ch}\Delta f + R_s/2\right]\right)$$
$$- \ln\left(2\alpha - jq f_1\left[n_{ch}\Delta f - R_s/2\right]\right)$$
$$+ \sum_{n_s=1}^{N_s-1} e^{-2\alpha L_{acc}^{n_s}} \Big[ \text{Ei}\left(L_{acc}^{n_s}\left[2\alpha - jq f_1\left[n_{ch}\Delta f + R_s/2\right]\right]\right)$$
$$- \text{Ei}\left(L_{acc}^{n_s}\left[2\alpha - jq f_1\left[n_{ch}\Delta f - R_s/2\right]\right]\right)\Big]$$
$$- \sum_{n_s=1}^{N_s} e^{-2\alpha L_{acc}^{n_s}} e^{-2\alpha L_{acc}^{n_s}} \Big[ \text{Ei}\left(L_{acc}^{n_s}\left[2\alpha - jq f_1\left[n_{ch}\Delta f + R_s/2\right]\right]\right)$$
$$- \text{Ei}\left(L_{acc}^{n_s}\left[2\alpha - jq f_1\left[n_{ch}\Delta f - R_s/2\right]\right]\right)\Big]\Big\}$$

**Eq. 18**

where 'Ei' is the exponential-integral function. The remaining single-dimensional integration needed to solve Eq. 16 could be easily and quickly carried out using any mathematical software. The result is then very accurate for all values of $N_s$ and fully accounts for link spans that can have arbitrarily different lengths. It can be generalized to spans having arbitrarily different fibers, but this further step is omitted.

However, we are interested in a simple closed-form approximation for $G_{corr}^{ex}(0)$ which provides insight into its basic fundamental parameter dependencies. To simplify the following derivation, we assume from now on that spans are all identical. Hence, the link function $\mu(f_1, f_2, 0)$ appearing in Eq. 17 can be re-written as:

$$\mu(f_1, f_2, 0) = \frac{1 - e^{-2\alpha L_s} e^{jq f_1 f_2 L_s}}{2\alpha - jq\, f_1 f_2} \sum_{n=0}^{N_s-1} e^{jn f_1 f_2 q L_s}$$
$$= \frac{1 - e^{-2\alpha L_s} e^{jq f_1 f_2 L_s}}{2\alpha - jq\, f_1 f_2} e^{jf_1 f_2 q(N_s-1)L_s/2} \frac{\sin\left(N_s f_1 f_2 q L_s/2\right)}{\sin\left(f_1 f_2 q L_s/2\right)}$$

**Eq. 19**

The fraction involving sine functions can be exactly expressed as a summation of $\sin(N_s[x - n\pi])/(x - n\pi)$ as follows:

$$\mu(f_1, f_2, 0) = \frac{1 - e^{-2\alpha L_s} e^{jq f_1 f_2 L_s}}{2\alpha - jq\, f_1 f_2} e^{jf_1 f_2 q(N_s-1)L_s/2} \cdot$$
$$\sum_{n=-\infty}^{\infty} \frac{\sin\left(N_s\left[f_1 f_2 q L_s/2 - n\pi\right]\right)}{f_1 f_2 q L_s/2 - n\pi}, \qquad N_s \text{ odd}$$

**Eq. 20**

Note the constraint that $N_s$ should be odd. A similar expression, involving two distinct summations, can be written for $N_s$ even. However, the final result of our approximations is independent of $N_s$ being even or odd. Therefore in the following, for the sake of notational compactness, we discuss the case of odd $N_s$ only and we also omit to explicitly re-write the constraint of $N_s$ being odd following each equation.

Each element of the summation in Eq. 20 has the form of a 'sinc' function: $\sin(N_s[x-n\pi])/(x-n\pi)$, whose main lobe width (null to null) vs. the integration variables product $f_1 f_2$, is: $W = 4\pi/(qN_s L_s)$. This shows that the main lobe width shrinks as $N_s$ goes up. As a result, for increasing $N_s$, each 'sinc' term in the summation of Eq. 20 tends to have a 'sampling' effect[4] vs. the slowly-varying factor present in the link function, i.e., vs.:

$$\frac{1 - e^{-2\alpha L_s} e^{jq f_1 f_2 L_s}}{2\alpha - jq \, f_1 f_2}$$

**Eq. 21**

Since we are interested in an approximation which is valid for large $N_s$, in this factor we can then replace $f_1 f_2$ with the location of the peak of the main lobes, which occur at: $f_1 f_2 = 2n\pi/(qL_s)$. As a result, we can write:

$$\mu(f_1, f_2, 0) \approx \mu^{(a)}(f_1, f_2, 0) = e^{jf_1 f_2 q(N_s-1)L_s/2} L_{\text{eff}} \cdot$$
$$\sum_{n=-\infty}^{\infty} \frac{\sin(N_s[f_1 f_2 qL_s/2 - n\pi])}{f_1 f_2 qL_s/2 - n\pi} \frac{1}{1 - jn\frac{\pi}{\alpha L_s}}$$

**Eq. 22**

where $\mu^{(a)}$ indicates an approximation of $\mu$ for large $N_s$. As a result:

$$\zeta_{n_{\text{ch}}}(f_1) \approx \zeta_{n_{\text{ch}}}^{(a)}(f_1) = \int_{n_{\text{ch}}\Delta f - R_s/2}^{n_{\text{ch}}\Delta f + R_s/2} \mu^{(a)}(f_1, f_2, 0) \, df_2$$

Apart from effectively sampling the slowly varying factor Eq. 21, the peak-wise nature of the summation in Eq. 20 also brings about another important effect. When carrying out the final integration in $f_1$ shown in Eq. 16, the following approximation can be used:

---

$$\int_{-R_s/2}^{+R_s/2} \left| \zeta_{n_{\text{ch}}}^{(a)}(f_1) \right|^2 df_1 = \int_{-R_s/2}^{+R_s/2} \left| \int_{n_{\text{ch}}\Delta f - R_s/2}^{n_{\text{ch}}\Delta f + R_s/2} \mu^{(a)}(f_1, f_2, 0) df_2 \right|^2 df_1$$

$$\approx \int_{-R_s/2}^{+R_s/2} \left| \int_{n_{\text{ch}}\Delta f - R_s/2}^{n_{\text{ch}}\Delta f + R_s/2} \mu_0^{(a)}(f_1, f_2, 0) df_2 \right|^2 df_1 +$$

$$\int_{-R_s/2}^{+R_s/2} \left| \sum_{\substack{n=-\infty \\ n\neq 0}}^{\infty} \int_{n_{\text{ch}}\Delta f - R_s/2}^{n_{\text{ch}}\Delta f + R_s/2} \mu_n^{(a)}(f_1, f_2, 0) df_2 \right|^2 df_1$$

**Eq. 23**

having defined:

$$\mu_n^{(a)}(f_1, f_2, 0) = \frac{\sin\left(N_s\left[\frac{f_1 f_2 qL_s}{2} - n\pi\right]\right)}{\frac{f_1 f_2 qL_s}{2} - n\pi} \frac{e^{jf_1 f_2 q(N_s-1)L_s/2} L_{\text{eff}}}{1 - jn\frac{\pi}{\alpha L_s}}$$

**Eq. 24**

This approximation is justified because the function $\left|\zeta_{n_{\text{ch}}}^{(a)}(f_1)\right|^2$ has a main peak, or main 'lobe', which is generated by $\mu_0^{(a)}(f_1, f_2, 0)$ around $f_1 = 0$. This peak, as we argue below, is separate from the other peaks of $\left|\zeta_{n_{\text{ch}}}^{(a)}(f_1)\right|^2$ which are generated by the terms $\mu_{n\neq 0}^{(a)}(f_1, f_2, 0)$. As a consequence, the peak at $f_1 = 0$ of $\left|\zeta_{n_{\text{ch}}}^{(a)}(f_1)\right|^2$ interacts only with itself within the $|\cdot|^2$ operator and can be extracted additively.

This feature of $\left|\zeta_{n_{\text{ch}}}^{(a)}(f_1)\right|^2$ is apparent in any numerical plot of it. It can be analytically justified as follows. The main lobe of the sinc-like factor $\mu_0^{(a)}(f_1, f_2, 0)$ extends (null to null) between:

$$\frac{2\pi}{qL_s}\left(n - \frac{1}{N_s}\right) \leq f_1 f_2 \leq \frac{2\pi}{qL_s}\left(n + \frac{1}{N_s}\right)$$

**Eq. 25**

For $n = 0$ such lobe falls within:

$$-\frac{1}{N_s}\frac{2\pi}{qL_s} \leq f_1 f_2 \leq \frac{1}{N_s}\frac{2\pi}{qL_s}$$

**Eq. 26**

Taking the lower limit of the integration range of $f_2$, at most such lobe can extend in $f_1$ symmetrically:

$$-\frac{1}{N_s}\frac{2\pi}{qL_s(n_{\text{ch}}\Delta f - R_s/2)} \leq f_1 \leq \frac{1}{N_s}\frac{2\pi}{qL_s(n_{\text{ch}}\Delta f - R_s/2)}$$

**Eq. 27**

Restarting from Eq. 25, the lobes generated by the terms for $n \neq 0$ show a null-to-null range that is instead (approximately):

$$\frac{2\pi}{qL_s\left(n_{\mathrm{ch}}\Delta f + R_s/2\right)}\left(n - \frac{1}{N_s}\right) \leq f_1 \leq \frac{2\pi}{qL_s\left(n_{\mathrm{ch}}\Delta f - R_s/2\right)}\left(n + \frac{1}{N_s}\right)$$

**Eq. 28**

This result is found from Eq. 25, by replacing $f_2$ with its upper integration limit to find the lower bound, and by its lower integration limit to find the upper bound.

Comparing Eq. 27 and Eq. 28 it is clear that, while the width in $f_1$ of the peak generated for $n = 0$ keeps shrinking for increasing $N_s$, the peak width for $n \neq 0$ tends to quickly become independent of $N_s$. Hence, separation of the lobe generated by $\mu_0^{(\mathrm{a})}\left(f_1, f_2, 0\right)$ from the others is ensured for large-enough $N_s$. In practice, comparing Eq. 27 and Eq. 28, it can be seen that $N_s = 4$ already ensures separation for Nyquist-WDM systems (i.e., $\Delta f = R_s$), at any value of $n_{\mathrm{ch}}$. Again, direct plots of $\left|\zeta_{n_{\mathrm{ch}}}^{(\mathrm{a})}\left(f_1\right)\right|^2$ confirm these results. Hence, from now on, we adopt Eq. 23 as our working approximation.

In the following, we first concentrate on the contribution in Eq. 23 due to the summation term for $n = 0$, and neglect all other contributions. Using this further approximation we will derive Eq. 2. We will then come back to the contributions for $n \neq 0$ to analyze and discuss them.

### A. The contribution for $n = 0$

The inner integral (in $f_2$) of the contribution to Eq. 23 for $n = 0$ can be solved analytically, yielding:

$$\int_{-R_s/2}^{+R_s/2}\left|\int_{n_{\mathrm{ch}}\Delta f - R_s/2}^{n_{\mathrm{ch}}\Delta f + R_s/2}\mu_0^{(\mathrm{a})}\left(f_1, f_2, 0\right)df_2\right|^2 df_1 =$$

$$L_{\mathrm{eff}}^2\int_{-R_s/2}^{+R_s/2}\left|\int_{n_{\mathrm{ch}}\Delta f - R_s/2}^{n_{\mathrm{ch}}\Delta f + R_s/2}\frac{\sin\left(\frac{1}{2}N_s f_1 f_2 qL_s\right)}{\frac{1}{2}f_1 f_2 qL_s}e^{jf_1 f_2 q(N_s-1)L_s/2}df_2\right|^2 df_1 =$$

$$L_{\mathrm{eff}}^2\int_{-R_s/2}^{+R_s/2}df_1\left|\frac{1}{qL_s f_1}\left[\,j\mathrm{cosint}\left(\tfrac{1}{2}\left[n_{\mathrm{ch}}\Delta f + R_s/2\right]qL_s f_1\right)\right.\right.$$

$$-j\mathrm{cosint}\left(\tfrac{1}{2}\left[n_{\mathrm{ch}}\Delta f - R_s/2\right]qL_s f_1\right)$$

$$+j\mathrm{cosint}\left(\left[N_s - \tfrac{1}{2}\right]\left[n_{\mathrm{ch}}\Delta f + R_s/2\right]qL_s f_1\right)$$

$$-j\mathrm{cosint}\left(\left[N_s - \tfrac{1}{2}\right]\left[n_{\mathrm{ch}}\Delta f - R_s/2\right]qL_s f_1\right)$$

$$+\mathrm{sinint}\left(\tfrac{1}{2}\left[n_{\mathrm{ch}}\Delta f + R_s/2\right]qL_s f_1\right)$$

$$-\mathrm{sinint}\left(\tfrac{1}{2}\left[n_{\mathrm{ch}}\Delta f - R_s/2\right]qL_s f_1\right)$$

$$-\mathrm{sinint}\left(\left[N_s - \tfrac{1}{2}\right]\left[n_{\mathrm{ch}}\Delta f - R_s/2\right]qL_s f_1\right)$$

$$\left.\left.+\mathrm{sinint}\left(\left[N_s - \tfrac{1}{2}\right]\left[n_{\mathrm{ch}}\Delta f + R_s/2\right]qL_s f_1\right)\right]\right|^2$$

**Eq. 29**

Unfortunately, we could not find an analytical solution for the following outer integral (in $f_1$). However, if the integration range over $f_1$ is extended to $\left[-\infty, \infty\right]$, then a simple closed-form overall result is found:

$$\int_{-R_s/2}^{+R_s/2}\left|\int_{n_{\mathrm{ch}}\Delta f - R_s/2}^{n_{\mathrm{ch}}\Delta f + R_s/2}\mu_0^{(\mathrm{a})}\left(f_1, f_2, 0\right)df_2\right|^2 df_1$$

$$\approx L_{\mathrm{eff}}^2\int_{-\infty}^{+\infty}\left|\int_{n_{\mathrm{ch}}\Delta f - R_s/2}^{n_{\mathrm{ch}}\Delta f + R_s/2}\frac{\sin\left(\frac{1}{2}N_s f_1 f_2 qL_s\right)}{\frac{1}{2}f_1 f_2 qL_s}e^{jf_1 f_2 q(N_s-1)L_s/2}df_2\right|^2 df_1$$

$$= \frac{4\pi N_s L_{\mathrm{eff}}^2 R_s}{qL_s}\left[1 + \left(n_{\mathrm{ch}}\frac{\Delta f}{R_s} - \frac{1}{2}\right)\log\left(\frac{n_{\mathrm{ch}}\Delta f/R_s - 1/2}{n_{\mathrm{ch}}\Delta f/R_s + 1/2}\right)\right]$$

**Eq. 30**

The error $\varepsilon$ incurred due to the extension of the $f_1$ integration range to $\left[-\infty, \infty\right]$ can be loosely upper-bounded as follows:

$$\varepsilon = 2L_{\mathrm{eff}}^2\int_{R_s/2}^{+\infty}\left|\int_{n_{\mathrm{ch}}\Delta f - R_s/2}^{n_{\mathrm{ch}}\Delta f + R_s/2}\frac{\sin\left(\frac{1}{2}N_s f_1 f_2 qL_s\right)}{\frac{1}{2}f_1 f_2 qL_s}e^{jf_1 f_2 q(N_s-1)L_s/2}df_2\right|^2 df_1 \leq$$

$$2L_{\mathrm{eff}}^2\int_{R_s/2}^{+\infty}\left|\int_{n_{\mathrm{ch}}\Delta f - R_s/2}^{n_{\mathrm{ch}}\Delta f + R_s/2}\frac{2}{f_1 f_2 qL_s}df_2\right|^2 df_1 = \frac{16L_{\mathrm{eff}}^2}{L_s^2 q^2 R_s}\log^2\left(\frac{n_{\mathrm{ch}}\Delta f/R_s + 1/2}{n_{\mathrm{ch}}\Delta f/R_s - 1/2}\right)$$

**Eq. 31**

Taking the ratio between Eq. 31 and Eq. 30, a loose upper bound to the relative error $\Delta$ due to the $f_1$ integration range extension can be found:

$$\Delta = \frac{1}{N_s} \cdot \frac{4\log^2\left(\dfrac{n_{ch}\Delta f / R_s + 1/2}{n_{ch}\Delta f / R_s - 1/2}\right)}{\pi L_s q R_s^2 \left[1 - \left(n_{ch}\dfrac{\Delta f}{R_s} - \dfrac{1}{2}\right)\log\left(\dfrac{n_{ch}\Delta f / R_s + 1/2}{n_{ch}\Delta f / R_s - 1/2}\right)\right]}$$

**Eq. 32**

Note that $\Delta$ goes down as $1/N_s$, which shows that the error due to this approximation asymptotically vanishes for large $N_s$. In any case, it is easy to check that the values of $\Delta$ that are found for typical values of system parameters are small or negligible. This allows to consider the $f_1$ integration range extension to $[-\infty, \infty]$ as a reliable approximation.

Substituting Eq. 30 into Eq. 16 we get:

$$G_{corr}^{ex}(0) \approx \sum_{n_{ch}\in\mathcal{N}_{ch}} \Phi \frac{80}{81} R_s^{-4} \gamma^2 P_{ch}^3 \frac{4\pi N_s L_{eff}^2 R_s}{qL_s} \cdot$$
$$\left[1 + \left(n_{ch}\frac{\Delta f}{R_s} - \frac{1}{2}\right)\log\left(\frac{n_{ch}\Delta f / R_s - 1/2}{n_{ch}\Delta f / R_s + 1/2}\right)\right]$$

**Eq. 33**

Eq. 33 can however be further simplified, by observing that:

$$\left[1 + \left(n_{ch}\frac{\Delta f}{R_s} - \frac{1}{2}\right)\log\left(\frac{n_{ch}\Delta f / R_s - 1/2}{n_{ch}\Delta f / R_s + 1/2}\right)\right] \approx \frac{R_s}{2n_{ch}\Delta f}$$

**Eq. 34**

The error of this approximation tends to vanish for increasing values of $n_{ch}$ and for increasing values of $\Delta f / R_s$. It is maximum for $n_{ch}=1$ and $\Delta f / R_s =1$, which are the minimum values for these quantities; there, it amounts to 0.45 dB. For $\Delta f / R_s =1$ and $n_{ch}=2$ it is down to 0.29 dB. Assuming to address a 10- or a 100-channel system, with $\Delta f / R_s =1$, the total error is 0.27 and 0.16 dB, respectively. These errors are modest and, given the much simpler analytical form of the right-hand-side of Eq. 34, vs. the left-hand side, the trade-off of accuracy vs. simplicity is arguably in its favor. Then substituting Eq. 34 into Eq. 33 yields:

$$G_{corr}^{ex}(0) \approx \sum_{n_{ch}\in\mathcal{N}_{ch}} \Phi \frac{40}{81} \frac{\gamma^2 P_{ch}^3 N_s L_{eff}^2}{R_s^2 \pi \beta_2 L_s \Delta f} \cdot \frac{1}{|n_{ch}|}$$

**Eq. 35**

Remembering the definition of the set $\mathcal{N}_{ch}$ given in Eq. 13, we have:

$$\sum_{n_{ch}\in\mathcal{N}_{ch}} \frac{1}{|n_{ch}|} = 2 \cdot \mathrm{HN}\left(\left[N_{ch}-1\right]/2\right)$$

**Eq. 36**

Substituting Eq. 36 into Eq. 35, Eq. 2 is found. The only residual difference is the presence in Eq. 2 of the average quantities $\overline{L_{eff}^2}$ and $\overline{L_s}$ rather than $L_{eff}^2$ and $L_s$. As mentioned in Sect. II, provided that each individual value of span length in the link stays within $\pm 15\%$ of the average, no significant loss of accuracy of Eq. 2 is found. The reason is that the main lobe of the $n$ =0 contribution in Eq. 23, from which the final result Eq. 35 is derived, is quite insensitive to fluctuations in the individual span lengths, at least within the indicated range.

### B. The contributions for $n \neq 0$

The contribution in Eq. 23 for $n \neq 0$ is:

$$\int_{-R_s/2}^{+R_s/2}\left|\sum_{\substack{n=-\infty\\n\neq 0}}^{\infty}\int_{n_{ch}\Delta f - R_s/2}^{n_{ch}\Delta f + R_s/2}\mu_n^{(a)}(f_1,f_2,0)df_2\right|^2 df_1 = L_{eff}^2$$

$$\int_{-R_s/2}^{+R_s/2}\left|\sum_{\substack{n=-\infty\\n\neq 0}}^{\infty}\int_{n_{ch}\Delta f - R_s/2}^{n_{ch}\Delta f + R_s/2}\frac{\sin\left(N_s\left[\dfrac{f_1 f_2 qL_s}{2} - n\pi\right]\right)}{\dfrac{f_1 f_2 qL_s}{2} - n\pi}\frac{e^{jf_1 f_2 q(N_s-1)L_s/2}L_{eff}}{1 - jn\dfrac{\pi}{\alpha L_s}}df_2\right|^2 df_1$$

**Eq. 37**

Note that this contribution is omitted altogether in Eq. 2. On the other hand, the quite compelling numerical validation of Eq. 2 in the wide variety of system configuration addressed in Figs. 1 - 6 indicate that, for large $N_s$, such contribution must be vanishing versus the one for $n = 0$.

In the following we try to show that this is the case, based on analytical results too. Unfortunately, we could not carry out the integrations in Eq. 37 in closed-form. We cannot rule out the possibility that such result can be found, but we decided to resort to the use of some further approximations to get to a closed-form result. This greater degree of approximation can be justified based on the fact that we are not actually interested in an accurate estimate of Eq. 37. Rather, we only want to characterize its general dependence on the main parameters, and especially on $N_s$.

First of all, we adopt the additive approximation for all terms in the integral for $n \neq 0$:

$$\int_{-R_s/2}^{+R_s/2}\left|\sum_{\substack{n=-\infty\\n\neq 0}}^{\infty}\int_{n_{ch}\Delta f - R_s/2}^{n_{ch}\Delta f + R_s/2}\mu_n^{(a)}(f_1,f_2,0)df_2\right|^2 df_1 \approx$$

$$\sum_{\substack{n=-\infty\\n\neq 0}}^{\infty}\int_{-R_s/2}^{+R_s/2}\left|\int_{n_{ch}\Delta f - R_s/2}^{n_{ch}\Delta f + R_s/2}\mu_n^{(a)}(f_1,f_2,0)df_2\right|^2 df_1 = \sum_{\substack{n=-\infty\\n\neq 0}}^{\infty}\int_{-R_s/2}^{+R_s/2}\left|\zeta_{n_{ch},n}^{(a)}(f_1)\right|^2 df_1$$

**Eq. 38**

having defined:

$$\left|\zeta_{n_{\mathrm{ch}},n}^{(\mathrm{a})}\left(f_1\right)\right|^2 = \left|\int_{n_{\mathrm{ch}}\Delta f - R_s/2}^{n_{\mathrm{ch}}\Delta f + R_s/2} \mu_n^{(\mathrm{a})}\left(f_1,f_2,0\right) df_2\right|^2$$

**Eq. 39**

We then remark that according to Eq. 25, the main lobe in the integrand function $\mu_n^{(\mathrm{a})}\left(f_1,f_2,0\right)$ has null-to-null width:

$$\frac{2\pi}{qL_s}\left(n-\frac{1}{N_s}\right) \le f_1 f_2 \le \frac{2\pi}{qL_s}\left(n+\frac{1}{N_s}\right)$$

We then observe that the center value of the integration range in $f_2$ is $n_{\mathrm{ch}}\Delta f$ and we assign such value to $f_2$ in the above equation. We also assume large $N_s$. Based on this we conclude that the center value of $f_1$ at which a peak should be observed in the function: $\left|\zeta_{n_{\mathrm{ch}},n}^{(\mathrm{a})}\left(f_1\right)\right|^2$ is approximately:

$$f_{1_{\mathrm{peak}}} = \frac{2\pi n}{qL_s n_{\mathrm{ch}}\Delta f}$$

**Eq. 40**

Numerical plots of $\left|\zeta_{n_{\mathrm{ch}},n}^{(\mathrm{a})}\left(f_1\right)\right|^2$ show this formula to be quite accurate in identifying the position of the peaks. We now calculate the value taken on by $\left|\zeta_{n_{\mathrm{ch}},n}^{(\mathrm{a})}\left(f_1\right)\right|^2$ at such peaks, that is we substitute $f_{1_{\mathrm{peak}}}$ into $\left|\zeta_{n_{\mathrm{ch}},n}^{(\mathrm{a})}\left(f_1\right)\right|^2$ and calculate the result. This step can be carried out in closed-form:

$$\left|\zeta_{n_{\mathrm{ch}},n}^{(\mathrm{a})}\left(f_{1,\mathrm{peak}}\right)\right|^2 = \left|\int_{n_{\mathrm{ch}}\Delta f - R_s/2}^{n_{\mathrm{ch}}\Delta f + R_s/2} \mu_n^{(\mathrm{a})}\left(f_{1,\mathrm{peak}},f_2,0\right) df_2\right|^2 =$$

$$= \left|\frac{L_{\mathrm{eff}}}{1 - j\frac{n\pi}{\alpha L_s}} \int_{n_{\mathrm{ch}}\Delta f - R_s/2}^{n_{\mathrm{ch}}\Delta f + R_s/2} \frac{\sin\left(N_s\left[\frac{n\pi}{n_{\mathrm{ch}}\Delta f}f_2 - n\pi\right]\right)}{\frac{n\pi}{n_{\mathrm{ch}}\Delta f}f_2 - n\pi} e^{j\frac{n\pi}{n_{\mathrm{ch}}\Delta f}f_2(N_s-1)} df_2\right|^2$$

$$= \left|\frac{L_{\mathrm{eff}}}{1 - j\frac{n\pi}{\alpha L_s}} \frac{n_{\mathrm{ch}}\Delta f}{n\pi}\left[\mathrm{sinint}\left(\left(2N_s-1\right)\frac{n\pi R_s}{2n_{\mathrm{ch}}\Delta f}\right) + \mathrm{sinint}\left(\frac{n\pi R_s}{2n_{\mathrm{ch}}\Delta f}\right)\right]\right|^2$$

**Eq. 41**

Then, we approximate $\left|\zeta_{n_{\mathrm{ch}},n}^{(\mathrm{a})}\left(f_1\right)\right|^2$ by assuming that its value is equal to $\left|\zeta_{n_{\mathrm{ch}},n}^{(\mathrm{a})}\left(f_{1,\mathrm{peak}}\right)\right|^2$ over the whole approximate extension of its main lobe, given by Eq. 28, and zero elsewhere:

$$\begin{cases} \left|\zeta_{n_{\mathrm{ch}},n}^{(\mathrm{a})}\left(f_1\right)\right|^2 \approx \left|\zeta_{n_{\mathrm{ch}},n}^{(\mathrm{a})}\left(f_{1,\mathrm{peak}}\right)\right|^2 \\[4pt] \frac{2\pi n}{qL_s\left(n_{\mathrm{ch}}\Delta f + R_s/2\right)}\left(n - \frac{1}{N_s}\right) \le f_1 \\[4pt] f_1 \le \frac{2\pi n}{qL_s\left(n_{\mathrm{ch}}\Delta f - R_s/2\right)}\left(n + \frac{1}{N_s}\right), \\[4pt] \left|\zeta_{n_{\mathrm{ch}},n}^{(\mathrm{a})}\left(f_1\right)\right|^2 \approx 0, \quad \text{elsewhere} \end{cases}$$

**Eq. 42**

Note that this approximation is rather crude, because $\left|\zeta_{n_{\mathrm{ch}},n}^{(\mathrm{a})}\left(f_1\right)\right|^2$ will not have the peak value throughout the extension of its main lobe. However, for now we keep it as is. We will come back later to this aspect.

Based on Eq. 42, the subsequent integration of $\left|\zeta_{n_{\mathrm{ch}},n}^{(\mathrm{a})}\left(f_1\right)\right|^2$ over $f_1$ is straightforward. Using these results, the whole contribution to Eq. 23 for $n \ne 0$ can then be approximated as:

$$\int_{-R_s/2}^{+R_s/2}\left|\sum_{\substack{n=-\infty\\n\ne0}}^{\infty}\int_{n_{\mathrm{ch}}\Delta f - R_s/2}^{n_{\mathrm{ch}}\Delta f + R_s/2} \mu_n^{(\mathrm{a})}\left(f_1,f_2,0\right) df_2\right|^2 df_1 \approx$$

$$\sum_{\substack{n=-\infty\\n\ne0}}^{\infty} \frac{2\pi L_{\mathrm{eff}}^2 R_s}{qL_s} \frac{1}{1 + \frac{n^2\pi^2}{\left(\alpha L_s\right)^2}} \frac{1}{|n|\pi^2} \frac{\left(1 + \frac{2n_{\mathrm{ch}}\Delta f}{|n|N_s R_s}\right)}{1 - \left(\frac{R_s}{2n_{\mathrm{ch}}\Delta f}\right)^2} \cdot$$

$$\left[\mathrm{sinint}\left((2N_s-1)\frac{|n|\pi R_s}{2n_{\mathrm{ch}}\Delta f}\right) + \mathrm{sinint}\left(\frac{|n|\pi R_s}{2n_{\mathrm{ch}}\Delta f}\right)\right]^2$$

**Eq. 43**

This closed-form formula shows some key features. First, in the RHS of Eq. 43 a factor approximately $1/|n|^3$ multiplies each term, which suggests that the strength of each successive peak goes down quite steeply vs. $n$, at least after the first sinint term has reached its asymptotic value $\pi/2$. In practice, the first few contributions typically already capture most of the contribution of the $n \ne 0$ terms, for large $N_s$.

Additionally, as $N_s$ goes up, the RHS of Eq. 43 tends to stop growing vs. $N_s$, after the first sinint term has reached its asymptotic value $\pi/2$. This is in contrast with Eq. 30 which shows that the contribution to Eq. 23 for $n = 0$ keeps on growing indefinitely as $N_s$. Hence, the indication is that the relative impact of the terms for $n \ne 0$ is asymptotically vanishing vs. $N_s$. This is coherent with the picture shown by Figs. 1 - 6, which is that $G_{\mathrm{corr}}$, which contains only the contribution for $n = 0$, underestimates the actual non-Gaussianity correction for low $N_s$, whereas for large $N_s$ it tends to be asymptotically accurate.

To find further confirmation of this interpretation, we did the following. We obtained a tentative $G_{corr}$, which we call $G'_{corr}$, which includes also the approximate contributions for $n \neq 0$. Specifically, using Eq. 31 and Eq. 43, we can write:

$$G'_{corr} \approx \sum_{n_{ch} \in N_{ch}} \Phi \frac{80}{81} R_s^{-4} \gamma^2 P_{ch}^3 \frac{4\pi L_{eff}^2 R_s}{qL_s}$$

$$\left\{ N_s \left[ 1 + \left( n_{ch} \frac{\Delta f}{R_s} - \frac{1}{2} \right) \log \left( \frac{n_{ch} \frac{\Delta f}{R_s} - \frac{1}{2}}{n_{ch} \frac{\Delta f}{R_s} + \frac{1}{2}} \right) \right] + \right.$$

$$+ \sum_{n=1}^{\infty} \frac{1}{1 + \frac{n^2 \pi^2}{(\alpha L_s)^2}} \frac{1}{2n\pi^2} \frac{\left( 1 + \frac{2n_{ch}\Delta f}{nN_s R_s} \right)}{1 - \left( \frac{R_s}{2n_{ch}\Delta f} \right)^2} \cdot$$

$$\left. \left[ \text{sinint} \left( (2N_s - 1) \frac{n\pi R_s}{2n_{ch}\Delta f} \right) + \text{sinint} \left( \frac{n\pi R_s}{2n_{ch}\Delta f} \right) \right]^2 \right\}$$

**Eq. 44**

In Fig. 8 (top) we plot XMCI for the same scenario addressed in Fig. 2 (top), i.e., for the PM-QPSK link over SMF with 15 channels. The black solid line is Eq. 44 which appears to be rather accurate for $N_s > 5$ but otherwise exhibits an overestimation of the non-Gaussianity correction for low $N_s$. However, as mentioned, Eq. 42 certainly leads to overestimating $\left| \zeta_{n_{ch},n}^{(a)} (f_1) \right|^2$. In fact, simply as a phenomenological and tentative exercise, by looking at the plots of the shape of the peaks in $\left| \zeta_{n_{ch},n}^{(a)} (f_1) \right|^2$, it can be argued that a factor $1/2$ should be approximately inserted to assign an 'effective value' which provides better accuracy for the subsequent integration over $f_1$:

$$\begin{cases} \left| \zeta_{n_{ch},n}^{(a)} (f_1) \right|^2 \approx \frac{1}{2} \cdot \left| \zeta_{n_{ch},n}^{(a)} (f_{1,peak}) \right|^2 \\ \frac{2\pi n}{qL_s (n_{ch}\Delta f + R_s/2)} \left( n - \frac{1}{N_s} \right) \leq f_1 \\ f_1 \leq \frac{2\pi n}{qL_s (n_{ch}\Delta f - R_s/2)} \left( n + \frac{1}{N_s} \right), \\ \left| \zeta_{n_{ch},n}^{(a)} (f_1) \right|^2 \approx 0, \quad \text{elsewhere} \end{cases}$$

**Eq. 45**

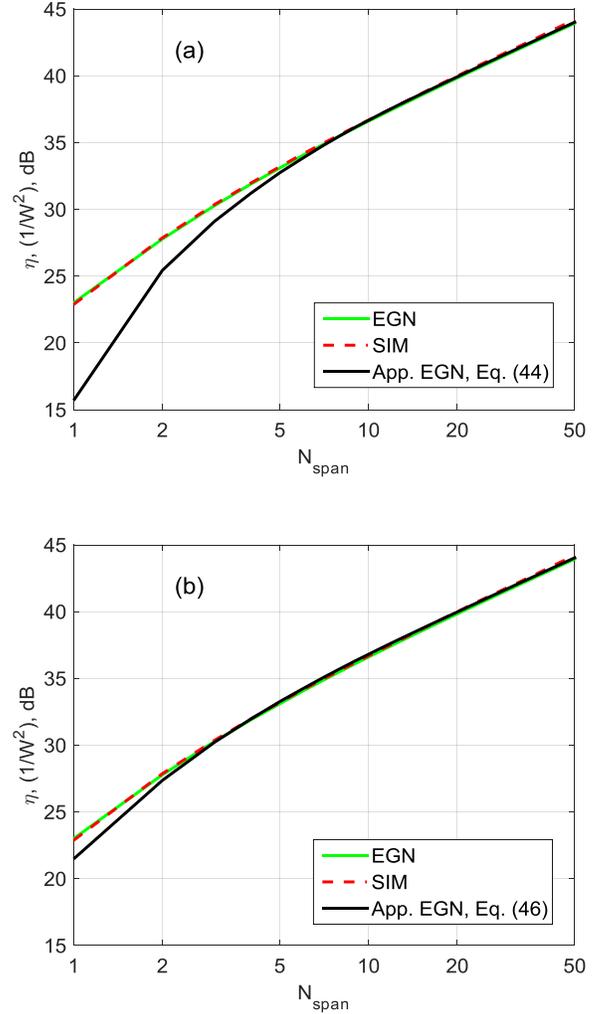

Fig. 8: Plot of the normalized combined cross- and multi-channel non-linearity noise power coefficient $\eta_{XMCI}$ affecting the center channel (in dB referred to $1 \cdot W^{-2}$), vs. number of spans in the link. Single-channel effects (SCI) are removed from all curves. System data: 15 PM-QPSK channels, symbol rate 32 GBaud, roll-off 0.05, SMF, span length 100 km, channel spacing 33.6 GHz. 'EGN' stands for EGN model result, 'SIM' for simulation result, 'App. EGN' for Eq. 6 where $G_{corr}$ is replaced by either (a) Eq. 44 or (b) Eq. 46.

Doing this, and deriving a new correction $G'_{corr}$:

$$G''_{corr} \approx \sum_{n_{ch} \in N_{ch}} \Phi \frac{80}{81} R_s^{-4} \gamma^2 P_{ch}^3 \frac{4\pi L_{eff}^2 R_s}{qL_s}$$

$$\left\{ N_s \left[ 1 + \left( n_{ch} \frac{\Delta f}{R_s} - \frac{1}{2} \right) \log \left( \frac{n_{ch} \dfrac{\Delta f}{R_s} - \dfrac{1}{2}}{n_{ch} \dfrac{\Delta f}{R_s} + \dfrac{1}{2}} \right) \right] + \right.$$

$$+ \sum_{n=1}^{\infty} \frac{1}{2} \frac{1}{1 + \dfrac{n^2\pi^2}{(\alpha L_s)^2}} \frac{1}{2n\pi^2} \frac{\left( 1 + \dfrac{2n_{ch}\Delta f}{nN_sR_s} \right)}{1 - \left( \dfrac{R_s}{2n_{ch}\Delta f} \right)} \cdot$$

$$\left. \left[ \mathrm{sinint}\left( (2N_s - 1) \frac{n\pi R_s}{2n_{ch}\Delta f} \right) + \mathrm{sinint}\left( \frac{n\pi R_s}{2n_{ch}\Delta f} \right) \right]^2 \right\}$$

**Eq. 46**

the much better result of Fig. 8 (bottom) is found.

Given the many approximations involved, we do not propose Eq. 46 as a reliable non-Gaussianity correction formula. However, we argue that its rather good coincidence with the EGN model and simulative curve provide further evidence that the $n \neq 0$ terms have the approximate behavior vs. $N_s$ that is shown by Eq. 43 and that this is why the asymptotic convergence of Eq. 2 was found to be good in the practical cases addressed in Figs. 1 - 6.